\documentclass[aps,preprint,nofootinbib]{revtex4}
\usepackage{amsfonts}
\usepackage{amsmath}
\usepackage{amssymb}
\usepackage{relsize}
\usepackage{graphicx}
\usepackage{pdfpages}
\usepackage{subfigure}
\usepackage{float}
\usepackage[caption = false]{subfig}
\usepackage{color}
\newcommand{\red}[1]{\textcolor{red}{#1}}
\newcommand{\blue}[1]{\textcolor{blue}{#1}}
\usepackage{multirow}

\def\correspondingauthor{\footnote{Corresponding author. E-mail address: liverts@phys.huji.ac.il}}
\begin{document}

\title{The collinear helium atom  and two-electron ions}

\author{Evgeny Z. Liverts\correspondingauthor{}}
\affiliation{Racah Institute of Physics, The Hebrew University, Jerusalem 91904, Israel}

\author{Rajmund Krivec}
\affiliation{Department of Theoretical Physics, J. Stefan Institute, 1000 Ljubljana, Slovenia}

\author{Nir Barnea}
\affiliation{Racah Institute of Physics, The Hebrew University, Jerusalem 91904, Israel}

\begin{abstract}
Collinear configurations of the helium-like atomic systems, relevant, e.g., for the quasifree mechanism of
the double photoionization of helium, are studied, parameterized by the single
scalar parameter $-1\leq \lambda\leq1$ (``collinear parameter") where
 $\lambda=0$ corresponds to the electron-nucleus ($\textbf{e-n}$)
coalescence and $\lambda=1$ corresponds to the
electron-electron ($\textbf{e-e}$) coalescence.
In general, $\lambda>0$ corresponds to the \textbf{n-e-e}
configuration, and $\lambda<0$ to the \textbf{e-n-e} configuration.
Simple mathematical representations of the expectation values of
the Dirac delta function relevant for the collinear configurations are
derived and calculated from fully three-body dynamics without approximation for the two-electron
atomic wave functions with  nuclear charge $1\leq Z\leq5$.
Simple formulas for calculating the expectation values of the
kinetic and potential energy operators in
collinear configurations are derived.
Unusual physical properties of the \textbf{n-e-e} collinear
configurations found for certain ranges of  $\lambda$ are presented.
The first few angular Fock coefficients for collinear configurations are derived as functions of $\lambda$.
Highly accurate model wave functions describing the ground states of the two-electron atoms with collinear arrangement of the particles are constructed. All results are illustrated by tables and figures.

\end{abstract}


\maketitle

\section{Introduction}\label{S0}

There are many scientific publications presenting numerical results of
expectation values of the Dirac delta operators $\langle \delta(\textbf{r}_1)\rangle$,
$\langle \delta(\textbf{r}_{12})\rangle\equiv \langle
\delta(\textbf{r}_1-\textbf{r}_2)\rangle$
and their products
$\langle\delta(\textbf{r}_1)\delta(\textbf{r}_{12})\rangle$
for the helium and the two-electron ions, where $\textbf{r}_1$, $\textbf{r}_2$ are the electron positions
relative to the nucleus (see, e.g., \cite{DRK,FR1,FR2,FR3} and references therein).
Quantum-mechanical applications of these expectation values can be found, e.g., in \cite{SUC,DAL,DRK}).
The two-electron wave functions (WF) at the two-particle coalescences,
$\psi(0,\textbf{r})$ and $\psi(\textbf{r},\textbf{r})$, had also been applied,
for example, in the description of the quasifree mechanism of double photoionization of helium (see, e.g., \cite{Druk,AM1,AM2}).
The collinear case of $\psi(\textbf{r},-\textbf{r})$  ($\lambda=-1$),
corresponding to the so called time-reversed double photoionization at
specific kinematics, had been studied in \cite{YA1,YA2}.

However, we have not found published results for $\langle
\delta(\textbf{r}_1+\textbf{r}_2)\rangle$ (corresponding to the boundary
value $\lambda=-1$), nor results for other values of $\lambda$.

Also, we have found a rather small number of articles devoted to the
particular case of the helium-like atomic systems in collinear configurations (see
\cite{TOL1,SIM1,SIM2,HAK}).
Moreover, to simplify the problem the authors considered a system of three particles constrained to move along a
straight line and interacting via the Coulomb forces \cite{TOL1}, or used the adiabatic approach \cite{SIM1,SIM2}.
Semiclassical calculations using the Herman-Kluk initial value treatment
had been performed in Ref. \cite{HAK} to
determine energies  of bound and resonance states of the collinear helium atom.

In contrast, in this work we employ the direct approach, calculating the non-relativistic WFs and the corresponding expectation values from two independent fully three-body methods for particles interacting via the Coulomb potentials, the Pekeris-like method (PLM) \cite{LEZ1, LEZ2}, and the correlation function hyperspherical harmonic method (CFHHM) \cite{HM1,HM2}.
Using proper three-body calculations together with certain
properties of the Dirac-delta function,
we were able to obtain interesting novel results, as well as to construct highly accurate parametrized analytical models of the three-body WFs in collinear configuration.

\section{Expectation values of the Dirac delta} \label{S1}


Let $\psi(\textbf{r}_1,\textbf{r}_2)$ represent the $S$-state solution of the Schr\"{o}dinger equation
\begin{equation}\label{1}
\left(T+V-E\right)\psi\left(\textbf{r}_1,\textbf{r}_2\right)=0,
\end{equation}
where $E$ denotes the non-relativistic electron energy of a
two-electron atom or ion with the infinitely massive nucleus of charge $Z$.
The atomic system of units will be used throughout the paper.
The kinetic energy operator reads
\begin{equation}\label{2}
T\equiv -\Delta/2,
\end{equation}
where $\Delta$ is the Laplacian.
The potential energy operator representing the interparticle Coulomb interactions is
\begin{equation}\label{3}
V\equiv -\frac{Z}{r_1}-\frac{Z}{r_2}+\frac{1}{r_{12}},
\end{equation}
where $r_1=|\textbf{r}_1|,~r_2=|\textbf{r}_2| $ and $r_{12}=|\textbf{r}_1-\textbf{r}_2|$ are the interparticle distances.

The \emph{collinear} arrangement of the particles (nucleus and two electrons) is defined by the relation
\begin{equation}\label{4}
\textbf{r}_1=\lambda \textbf{r}_2,
\end{equation}
where $\lambda\in [-1,1]$ is a scalar parameter (at least for the $S$-states treated here).
Clearly $\lambda=0$ corresponds to the \emph{electron-nucleus coalescence}, and $\lambda=1$ to the \emph{electron-electron coalescence}.
The boundary value $\lambda=-1$ corresponds to the collinear \textbf{e-n-e} configuration with the same distances of both electrons from the nucleus.
In general, $0 < \lambda \leq 1$ corresponds to the collinear arrangement of
the form \textbf{n-e-e} where both electrons
are on the same side of the nucleus.
Accordingly, $-1\leq\lambda < 0$ corresponds to the collinear arrangement
of the form \textbf{e-n-e} where the electrons are on the opposite sides of
the nucleus.
The absolute value $|\lambda|$ measures the ratio of the distances of the electrons from the nucleus.

Denoting for simplicity $\delta^3(\textbf{r})\equiv\delta(\textbf{r})$, and using the well-known properties of the Dirac-delta function, one obtains
\begin{equation}\label{5}
\langle\delta(\textbf{r}_1-\lambda \textbf{r}_2)\rangle\equiv\frac{1}{N}\int d^3\textbf{r}_2\int d^3\textbf{r}_1
\delta(\textbf{r}_1-\lambda \textbf{r}_2)\left|\psi(\textbf{r}_1,\textbf{r}_2)\right|^2=
\frac{1}{N}\int d^3\textbf{r}_2\left|\psi(\lambda \textbf{r}_2,\textbf{r}_2)\right|^2,
\end{equation}
where the normalization integral is
\begin{equation}\label{6}
N=\int d^3\textbf{r}_2\int d^3\textbf{r}_1\left|\psi(\textbf{r}_1,\textbf{r}_2)\right|^2.
\end{equation}
Clearly $\psi(\lambda \textbf{r}_2,\textbf{r}_2)$ represents
the WF describing the \emph{collinear arrangement} of the particles.

Taking into account the relationship
\begin{equation}\label{7}
\langle\delta(\textbf{r}_1)\delta(\textbf{r}_2)\rangle\equiv\
\frac{1}{N}\int d^3\textbf{r}_2\int d^3\textbf{r}_1
\delta(\textbf{r}_1)\delta(\textbf{r}_2)\left|\psi(\textbf{r}_1,\textbf{r}_2)\right|^2=\frac{1}{N}\psi^2(0,0),
\end{equation}
we can rewrite Eq. (\ref{5}) in the form
\begin{equation}\label{8}
\langle\delta(\textbf{r}_1-\lambda \textbf{r}_2)\rangle=\langle\delta(\textbf{r}_1)\delta(\textbf{r}_2)\rangle
\int d^3\textbf{r}_2 \left|\tilde{\psi}(\lambda \textbf{r}_2,\textbf{r}_2)\right|^2,
\end{equation}
where $\tilde{\psi}(\lambda \textbf{r}_2,\textbf{r}_2)=\psi(\lambda
\textbf{r}_2,\textbf{r}_2)/\psi(0,0)$ is the collinear WF
normalized such that $\tilde{\psi}(0,0)=1$.


It is well known that for the $S$-states the two-electron wave function $\psi(\textbf{r}_1,\textbf{r}_2)$ reduces to the function of only three internal coordinates (see, e.g., \cite{AB1}),
the simplest and most natural choice of which is the
set of the interparticle distances $r_1,~r_2$ and $r_{12}$.
This reduces the solution $\psi(\textbf{r}_1,\textbf{r}_2)$ of the
Schr\"{o}dinger equation (\ref{1}) to the form $\Phi(r_1,r_2,r_{12})$.
However, in our case a more convenient coordinate system is $\left\{ r_1,r_2,\theta\right\}$, where $\theta$ is the angle between the radius-vectors $\textbf{r}_1$ and $\textbf{r}_2$ of the electrons.
Let us denote the corresponding $S$-state WF as $\Psi(r_1,r_2,\theta)\equiv \Phi(r_1,r_2,r_{12})$.
The volume element in the $\left\{ r_1,r_2,\theta\right\}$ coordinates is
\begin{equation}\label{9}
d^3\textbf{r}_2d^3\textbf{r}_1=8\pi^2r_1^2 r_2^2  dr_1 dr_2 \sin \theta
d\theta, ~~~~~~~~~~ \theta\in[0,\pi].
\end{equation}
It can be shown that the Dirac delta for the $\left\{r_1,r_2,\theta\right\}$ coordinate system is similar to the one in the spherical coordinates with azimuthal symmetry, whence
\begin{equation}\label{10}
\delta\left(\textbf{r}_1-\lambda \textbf{r}_2\right)=\frac{1}{2\pi r_1^2 \sin \theta_1}
\delta(r_1-\left|\lambda\right|
r_2)\delta(\theta_1-\theta_2),~~~~~~~~(\theta_2=0,\pi)
\end{equation}
where $\theta_1$ and $\theta_2$ are the polar angles of the radius-vectors $\textbf{r}_1$ and $\textbf{r}_2$, respectively.
Inserting representations (\ref{9}) and (\ref{10}) into the RHS of Eq. (\ref{5}), one obtains
\begin{equation}\label{11}
\left\langle\delta\left(\textbf{r}_1-\lambda \textbf{r}_2\right)\right\rangle=\frac{4\pi}{N}\int_0^\infty
\left|\Psi\left(\left|\lambda\right|r_2,r_2,\theta_2\right)\right|^2r_2^2 dr_2=
\frac{4\pi}{N}\int_0^\infty \left|\Phi\left(\left|\lambda\right|r_2,r_2,(1-\lambda)r_2\right)\right|^2r_2^2 dr_2,
\end{equation}
where the angle $\theta_2=0$ corresponds to the collinear
configuration \textbf{n-e-e} ($\lambda>0$), whereas
$\theta_2=\pi$ corresponds to the collinear configuration \textbf{e-n-e}
($\lambda<0$).

It is seen that according to Eq.\ (\ref{5}) in the general case, and
according to Eq.\ (\ref{11}) for the $S$-state, the expectation value
$\left\langle\delta\left(\textbf{r}_1-\lambda\textbf{r}_2\right)\right\rangle$
reduces to the expectation value over the \emph{collinear} WF, $\Phi\left(\left|\lambda\right|r,r,(1-\lambda)r\right)$,
where $r$ is the distance between the nucleus and the electron most distant from it.
The collinear WF is therefore parameterized by a single scalar parameter $\lambda$.

\begin{figure}
\centering
\caption{The $\lambda$-normalized expectation values
$\left\langle\delta\left(\textbf{r}_1-\lambda
\textbf{r}_2\right)\right\rangle$, given by Eq.\ (\ref{11}), as functions of the \emph{collinear} parameter
$\lambda$ for the two-electron atomic systems considered. The solid line (black online) corresponds to the asymptotic two-electron ion with $Z\rightarrow \infty$, given by Eq.\ (\ref{11b}).}
\includegraphics[width=6.5in]{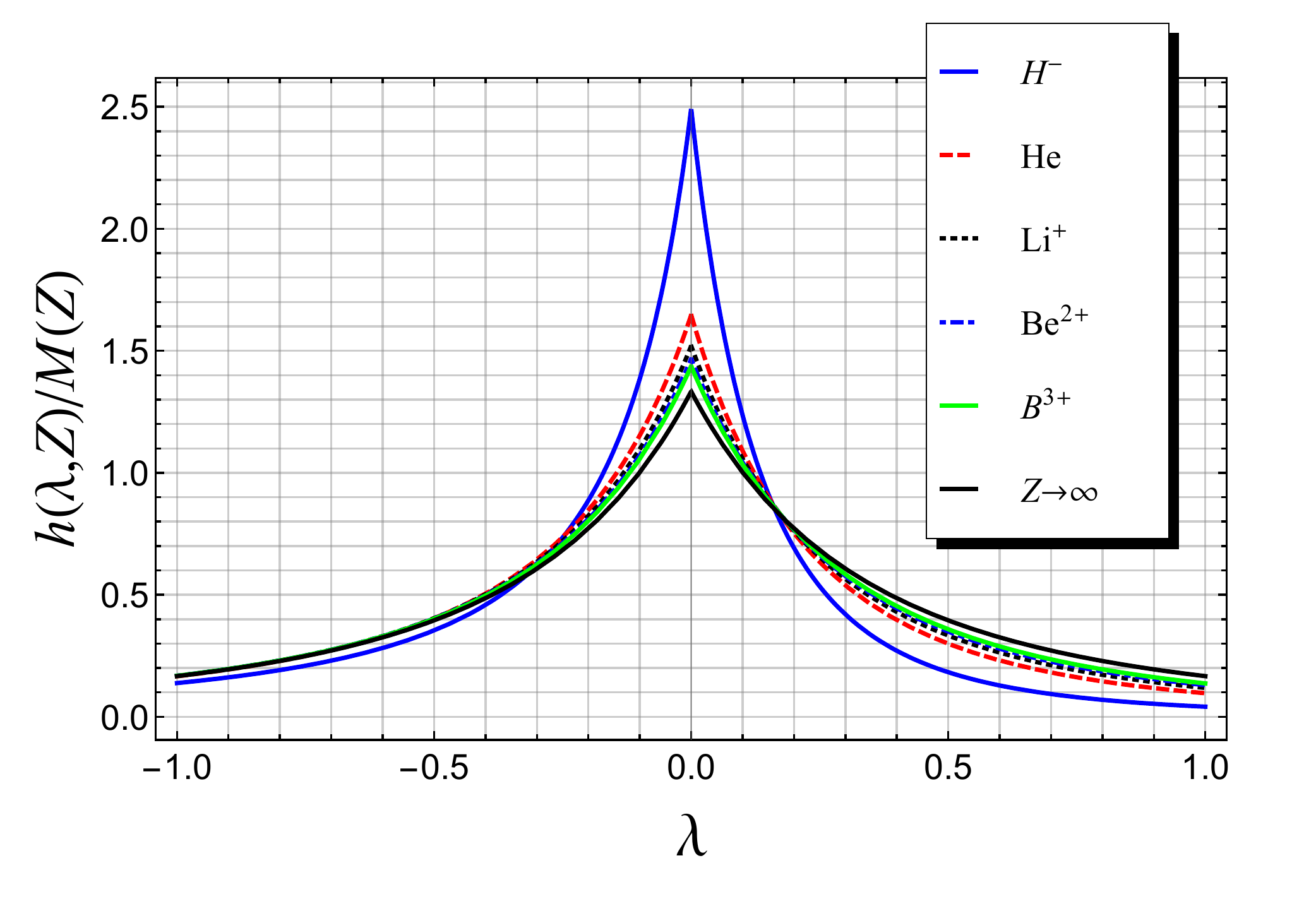}
\label{F1}
\end{figure}

Using the Pekeris-like three-body method \cite{LEZ1,LEZ2} we have calculated the
expectation value
$\mathrm{h}(\lambda,Z)\equiv\left\langle\delta\left(\textbf{r}_1-\lambda
\textbf{r}_2\right)\right\rangle$ for the ground states of the two-electron
atomic systems with $1\leq Z\leq5$.
Using the normalization parameter over $\lambda$, defined as
\begin{equation}\label{11a}
M(Z)=\int_{-1}^1 \mathrm{h}(\lambda,Z)d \lambda,
\end{equation}
we can present the plots of $\Gamma_Z(\lambda)\equiv\mathrm{h}(\lambda,Z)/M(Z)$ for all considered $Z$ on a single figure (see Fig. \ \ref{F1}).
The derivative $d\mathrm{h}(\lambda,Z)/d \lambda$ is obviously singular at $\lambda=0$.
In particular, for the helium atom we obtain $\underset{\lambda\rightarrow 0^+}{\lim} d\mathrm{h}(\lambda,2)/d \lambda\simeq -7.92 $, whereas
 $\underset{\lambda\rightarrow 0^-}{\lim} d\mathrm{h}(\lambda,2)/d \lambda\simeq 6.96$.

The parts of the curves in Fig.\ \ref{F1} for $\lambda<0$ and $\lambda>0$ are asymmetric with respect to $\lambda=0$.
However, they have two readily apparent properties: (i) rapid convergence with increasing $Z$, and (ii) the tendency toward symmetry with increasing $Z$.
This indicates the possibility of existence
of an analytic expression valid asymptotically ($Z\rightarrow \infty$) which is symmetric with respect to the sign of the \emph{collinear} parameter $\lambda$.
To find such an asymptotic form,
let us suppose that for large enough $Z$ we can neglect the electron-electron interaction in comparison with the electron-nucleus interaction in the Schr\"{o}dinger equation (\ref{1})-(\ref{3}). It is well-known that the corresponding ground state solution is of the form $\Phi_\infty\sim\exp\left[-Z (r_1+r_2)\right]=\exp\left[-Z r (1+|\lambda|)\right]$.
Inserting this solution into Eq.(\ref{6}) we obtain the normalization parameter $N_\infty=\pi^2/Z^6$.
Subsequent substitution of $\Phi_\infty$ and $N_\infty$ into Eqs.(\ref{11}) and (\ref{11a}) yields $h(\lambda,Z)=Z^3/\pi(1+|\lambda|)^3$ and $M(Z)=3Z^3/4\pi$, respectively,
resulting in the asymptotic expression
 \begin{equation}\label{11b}
\Gamma_\infty(\lambda)=\frac{4}{3\left(1+|\lambda|\right)^3}.
\end{equation}
The curve (\ref{11b}) is shown in Fig. \ref{F1} by a solid line (black online) that very accurately agrees with the asymptotic behavior. For additional verification we calculated, using the PLM, the boundary values at $Z=100$:
$\Gamma_{100}(0)\simeq 1.3380$, $\Gamma_{100}(-1)\simeq 0.16676$ and $\Gamma_{100}(1)\simeq 0.16513$.
The corresponding values following from Eq.(\ref{11b}) are:
$\Gamma_\infty(0)\simeq 1.3333$, $\Gamma_\infty(-1)=\Gamma(1,\infty)=\simeq 0.16667$,
confirming the validity of the asymptotic expression (\ref{11b}).

Table \ref{T1} presents the values of $h(\lambda,Z)$
for the \emph{collinear} parameter $\lambda=-1,-0.5,0,0.5,1$,
as well as for some specific values of $\lambda$
whose physical meaning will be clarified in the next section.
The values of normalization $M(Z)$ are presented as well.
To estimate the general accuracy of our calculations, we have presented the ground state energies $E$ calculated by the Pekeris-like method \cite{LEZ2} with number of shells $\Omega=25$.
For comparison, the results of the more accurate calculations are available \cite{DRK,FR1,FR2,FR3}, they can be obtained by replacing our last digit by the adjacent digit in square brackets.

\section{Expectation values of the product of Dirac delta and Hamiltonian} \label{S2}

\begin{figure}
\caption{Expectation values of $T=-\Delta/2$ (circles) and
$V=r^{-1}\left[-Z-Z/\lambda+1/(1-\lambda)\right]$ (triangles)
for the helium atom in the \emph{collinear} $1S$ state.
$r$ is the distance between the nucleus and one of the electrons, and
$|\lambda|r$ is the distance between the nucleus and the other
electron. The (\textbf{e-n-e}) configuration corresponds to $\lambda <
0$, the (\textbf{n-e-e}) configuration - to $\lambda > 0$.}
\includegraphics[width=6.0in]{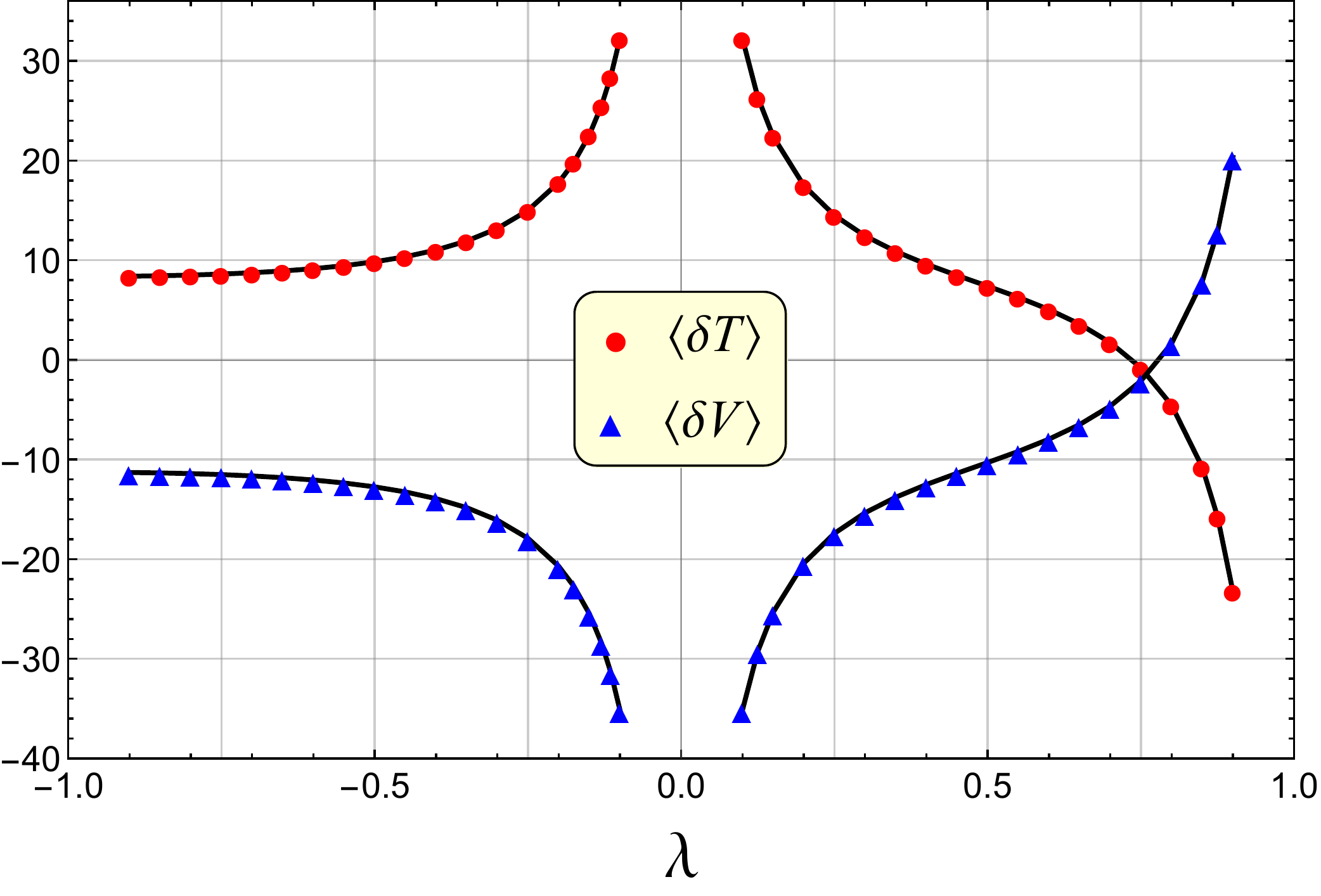}
\label{F2}
\end{figure}

Let us multiply the Schr\"{o}dinger equation (\ref{1}) on the left by
$\delta\left(\textbf{r}_1-\lambda \textbf{r}_2\right)\psi(\textbf{r}_1,\textbf{r}_2)$
and integrate both sides over the whole space. This yields
\begin{equation}\label{12}
\int d^3\textbf{r}_2\int d^3\textbf{r}_1\delta(\textbf{r}_1-\lambda \textbf{r}_2)\psi(\textbf{r}_1,\textbf{r}_2)(T+V)\psi(\textbf{r}_1,\textbf{r}_2)=
E\int d^3\textbf{r}_2\int d^3\textbf{r}_1\delta(\textbf{r}_1-\lambda \textbf{r}_2)\psi^2(\textbf{r}_1,\textbf{r}_2).
\end{equation}
Dividing both sides of Eq.\ (\ref{12}) by the RHS integral and simplifying, one obtains the relation
\begin{equation}\label{13}
\langle \delta T\rangle+\langle \delta V\rangle=E,
\end{equation}
where the term associated with expectation value of the kinetic energy operator in the $1S$ \emph{collinear} configuration is
\begin{equation}\label{14}
 \left\langle\delta T\right\rangle
  =-\frac{2 \pi}{N \mathrm{h}(\lambda,Z)}
  \int_0^\infty \Psi\left(\left|\lambda\right|r_2,r_2,\theta_2\right)\,
  \left[\Delta(r_1,r_2,\theta_1)\,\Psi\left( r_1,r_2,\theta_1\right)\right]_{\theta_1=\theta_2}^{r_1=|\lambda|r_2}\, r_2^2 d r_2,
\end{equation}
and $\mathrm{h}(\lambda,Z)$ is given by Eq.\ (\ref{11}).
Here $\theta_2=0$ for $\lambda >0$, and $\theta_2=\pi$ for $\lambda <0$.
In the derivation of Eq.\ (\ref{14}) we used representations
(\ref{9}) and (\ref{10}).
In the $\left\{r_1,r_2,\theta\right\}$ coordinate system the Laplacian is of the form (see, e.g., \cite{AB2})
\begin{equation}\label{15}
\Delta(r_1,r_2,\theta)=r_1^{-2}\frac{\partial}{\partial r_1}r_1^2\frac{\partial}{\partial r_1}+
r_2^{-2}\frac{\partial}{\partial r_2}r_2^2\frac{\partial}{\partial r_2}+\left(\frac{1}{r_1^2}+\frac{1}{r_2^2}\right)
(\sin\theta)^{-1}\frac{\partial}{\partial\theta}\sin \theta\frac{\partial}{\partial\theta}.
\end{equation}
For the term associated with the expectation value of the potential energy operator in the $1S$ \emph{collinear} configuration, we easily obtain:
\begin{equation}\label{16}
\langle \delta V\rangle
=\frac{4 \pi}{N \mathrm{h}(\lambda,Z)}\left(-Z-\frac{Z}{\lambda}+\frac{1}{1-\lambda}\right)
\int_0^\infty \Psi^2\left(\left|\lambda\right|r_2,r_2,\theta_2\right)r_2 dr_2.
\end{equation}
where we used the relation $r_{12}=r_2(1-\lambda)$ corresponding to the \emph{collinear} configuration.
The factor in parentheses enables us to conclude that the term $\langle \delta V\rangle$ vanishes when the parameter $\lambda$ takes the value
\begin{equation}\label{17}
\lambda_{v0}=\frac{\sqrt{1+4Z^2}-1}{2Z}.
\end{equation}
Using the Pekeris-like method \cite{LEZ1,LEZ2} we have calculated the
expectation values $\langle \delta V\rangle$ and $\langle \delta T\rangle$
for the ground state of the helium-like atomic systems with $1\leq Z\leq 5$.
The dependence of the expectation values mentioned above upon the
\emph{collinear} parameter $\lambda$ is displayed in Fig. \ref{F2} for the ground state of helium.
The plots for the two-electron ions are similar.
It can be seen that the plots in Fig. \ref{F2} exhibit some
characteristic points with peculiar behavior. One of them is defined by Eq. (\ref{17}).
The others are as follows:
\renewcommand{\labelenumi}{\theenumi)}
\begin{enumerate}
\item  the point $\lambda_{t0}>0$ at which $\left\langle \delta T \right\rangle=0$;

\item the point $\lambda_{2}>0$ at which $\left\langle \delta V
\right\rangle/\left\langle \delta T \right\rangle=-2$,
corresponding to the virial theorem for
the Coulomb interactions;

\item the crossing point $\lambda_{cr}>0$ when $\left\langle \delta V \right\rangle=\left\langle \delta T \right\rangle$;

\item the inflection point $\lambda_{v2}>0$ at which $d^2\left\langle \delta V \right\rangle/d \lambda^2=0$;

\item the inflection point $\lambda_{t2}>0$ at which $d^2\left\langle \delta T \right\rangle/d \lambda^2=0$;

\item the boundary point $\lambda=-1$.
\end{enumerate}

The characteristic points and some associated functions are listed in Table \ref{T2} for the helium-like
atomic systems under consideration.
We would like to highlight the following points:

\renewcommand{\theenumi}{\roman{enumi}}
\renewcommand{\labelenumi}{\theenumi)}
\begin{enumerate}

\item the most important and interesting features of the expectation values $\left\langle \delta V\right\rangle$ and  $\left\langle \delta T\right\rangle$ as functions of the parameter $\lambda$ are related to the collinear configuration \textbf{n-e-e} corresponding to $\lambda>0$;

\item the inflection points $\lambda_{v2}$ and $\lambda_{t2}$ (the latter not listed)
for
$\left\langle \delta V \right\rangle$ and  $\left\langle \delta V
\right\rangle$
coincide to at least four significant digits for each two-electron system;

\item interpreting $\mathrm{h}(\lambda,Z)/M(Z)$ (see Fig. \ref{F1}) as a function characterizing the probability
of formation of the collinear configuration with given $\lambda$ among all possible $\lambda\in [-1,1]$, we obtain that
$\mathrm{h}(\lambda_2,Z)/M(Z)$ gives the maximum value for $Z=2$.
In other words, the probability of the $\lambda_2$ collinear configuration
which satisfies the virial theorem
($\left\langle \delta V \right\rangle/\left\langle \delta T
\right\rangle=-2$) for the helium atom is higher than that for the two-electron ions, both negative and positive;

\item for the helium atom as well as for
each two-electron ion there are points $\lambda_{v0}>0$ and
$\lambda_{t0}>0$ at which the curves $\left\langle \delta V \right\rangle$
and  $\left\langle \delta T \right\rangle$  change
sign, i.e., $\left\langle \delta V \right\rangle=0$ for $\lambda=\lambda_{v0}$, and $\left\langle \delta T \right\rangle=0$ for $\lambda=\lambda_{t0}$.
Most importantly, $\lambda_{t0}<\lambda_{v0}$, and the derivative
$d\left\langle \delta T \right\rangle/d\lambda<0 $ at
$\lambda=\lambda_{t0}$, whereas $d\left\langle \delta V
\right\rangle/d\lambda>0 $ at $\lambda=\lambda_{v0}$.
It follows that for each two-electron atomic system there exists a region
$\lambda_{t0}\leq\lambda\leq\lambda_{v0}$ where both $\left\langle
\delta V \right\rangle$ and  $\left\langle \delta T \right\rangle$ are
negative.
Obviously the point $\lambda_{cr}$ is inside this region (see Fig. \ref{F2}).
\end{enumerate}
We are currently studying the applications of these results to the direct and time-reversed double photoionization, as well as examining possible applications of the unusual behavior of the expectation values
$\left \langle \delta V \right \rangle$ and $\left \langle \delta T \right \rangle$  in the $\lambda_{t0}< \lambda <\lambda_{v0}$ interval, as
we have reason to presume that it has special physical significance.

\section{The Fock expansion} \label{S3}

The behavior of the two-electron atomic WF, $\Phi(r_1,r_2,r_{12})$ in
the vicinity of the nucleus (located at the origin) is determined by the Fock expansion \cite{FOCK}
\begin{equation}\label{19}
\tilde{\Phi}(r_1,r_2,r_{12})\equiv\Phi(r_1,r_2,r_{12})/\Phi(0,0,0)=\sum_{k=0}^\infty R^k\sum_{p=0}^{[k/2]}\phi_{k,p}(\alpha,\theta)\ln^p R,
\end{equation}
where the hyperspherical coordinates $R,~\alpha$ and $\theta$ are defined as follows:
\begin{equation}\label{20}
R=\sqrt{r_1^2+r_2^2},~~~~\alpha=2\arctan\left(\frac{r_2}{r_1}\right),~~~~\theta=\arccos\left(\frac{r_1^2+r_2^2-r_{12}^2}{2 r_1 r_2}\right).
\end{equation}
The explicit form of the angular Fock coefficients (AFC) $\phi_{k,p}(\alpha,\theta)$ for low orders $k$ can be found in Ref. \cite{LEZ3} (see also Refs. \cite{AB1, AB2, AB3}).
Clearly $\phi_{0,0}=1$ for the representation (\ref{19}).
The other AFCs are:
\begin{equation}\label{21}
\phi_{1,0}=-Z\sqrt{1+\sin \alpha}+\frac{1}{2}\sqrt{1-\sin \alpha \cos \theta}=
\frac{1}{R}\left[-Z(r_1+r_2)+\frac{1}{2}r_{12}\right],
\end{equation}
\begin{equation}\label{22}
\phi_{2,1}=-Z\left(\frac{\pi-2}{3\pi}\right)\sin \alpha \cos \theta=
-Z\left(\frac{\pi-2}{3\pi}\right)\left(\frac{r_1^2+r_2^2-r_{12}^2}{R^2}\right),
\end{equation}
and
\begin{eqnarray}\label{23}
\phi_{3,1}=Z\left(\frac{\pi-2}{36\pi}\right)
\left[6Z\sin \alpha \cos \theta\sqrt{1+\sin \alpha}-(1+5\sin \alpha\cos \theta)\sqrt{1-\sin \alpha \cos \theta}\right]=
~\nonumber~~\\
=Z\left(\frac{\pi-2}{36\pi}\right)
\left\{6Z\frac{(r_1^2+r_2^2-r_{12}^2)(r_1+r_2)}{R^3}-\left[1+\frac{5r_{12}(r_1^2+r_2^2-r_{12}^2)}{R^3}
\right]\right\}.~~~~~~~~~~~~~
\end{eqnarray}
The AFC $\phi_{3,0}(\alpha,\theta)$ is  given in Ref. \cite{LEZ3} only partially, whereas the specific expression for $\phi_{2,0}(\alpha,\theta)$ is given by Eq. (22) in the same reference.

For the \emph{collinear} arrangement of the particles defined by the relations
\begin{equation}\label{24}
r_1=\lambda r,~~~~~~~~~~~~r_2=r,~~~~~~~~~~~~r_{12}=(1-\lambda)r,
\end{equation}
where $r=\textrm{max}\left\{r_1,r_2\right\}$, the Fock expansion (\ref{19}) becomes
\begin{equation}\label{25}
\tilde{\Phi}(|\lambda|r,r,(1-\lambda)r)\underset{r\rightarrow 0}{=}
1+\eta_\lambda r+\zeta_\lambda r^2 \ln r+\xi_\lambda r^2+\gamma_\lambda r^3\ln r+c_\lambda^{(30)}r^3+O(r^4),~~
\end{equation}
where the coefficients are:
\begin{eqnarray}
\eta_\lambda&=&-Z\left(1+|\lambda|\right)+\frac{1-\lambda}{2},\label{26}\\
\zeta_\lambda &=&-\frac{2 Z \lambda(\pi-2)}{3\pi},\label{27}\\
\gamma_\lambda &=&\frac{Z(\pi-2)}{36\pi}\left\{12 Z\, \lambda\,|\lambda|+\lambda\left[\lambda(\lambda+9)+12 Z-9\right]-1\right\}.\label{28}
\end{eqnarray}
We cannot calculate the coefficient $c_\lambda^{(30)}$ because the explicit form of the AFC $\phi_{3,0}(\alpha,\theta)$  has not been derived in a final form. However, we can do that for the coefficient $\xi_\lambda$. It is not a simple problem, because any one of the explicit forms of the AFC $\phi_{2,0}(\alpha,\theta)$ represents a quite complicated expression (see, e.g., \cite{AB3}, \cite{FRY}, \cite{MYERS} or \cite{LEZ3}). Thus, taking some nontrivial limits, we finally obtain:
\begin{equation}\label{29}
\xi_\lambda=(1+\lambda^2)\left\{\frac{1-2E}{12}+Z\left[\omega^{(\pm)}(\alpha)\pm a_{21} \sin \alpha\right]
+Z^2\left(\frac{1}{3}+\frac{1}{2}\sin \alpha\right)
\right\},
\end{equation}
where
\begin{equation}\label{32}
\alpha=2\arctan |\lambda|,
\end{equation}
\begin{eqnarray}\label{30}
\kern-2pt\omega^{(+)}(\alpha)\kern-2pt &=& \kern-2pt\frac{\cos\alpha}{6\pi}
\left\{2\alpha-3\pi+\pi\ln[2(1-\sin \alpha)\sec\alpha] - \pi\tan\alpha\left(\ln[2(1+\cos\alpha)]-1\right) - H(\alpha)\right\}\kern-2pt,\nonumber\\
&&\kern180pt\left(0\leq\lambda\leq 1\right),
\end{eqnarray}
\begin{eqnarray}\label{31}
\omega^{(-)}(\alpha)\kern-2pt &=& \kern-2pt-\frac{2}{3}+\frac{\cos \alpha}{6\pi}
\left\{\pi-2\alpha+\pi\ln(1+\sec \alpha)+ \pi \tan \alpha \left(\ln[4(1+\sin \alpha)]-3\right)+H(\alpha)\right\}\kern-1pt,\nonumber\\
&&\kern172.5pt\left(-1\leq\lambda \leq0\right),
\end{eqnarray}
and the function $H(\alpha)$ is given by
\begin{equation}\label{33}
H(\alpha)=-2\tan \alpha~_3F_2\left(\frac{1}{2},\frac{1}{2},1;\frac{3}{2},\frac{3}{2};-\tan^2\alpha\right)=
i\left[\textrm{Li}_2(i\tan \alpha)-\textrm{Li}_2(-i \tan \alpha)\right],\textbf{}
\end{equation}
where $_3F_2(...)$ is the hypergeometric function, and $\textrm{Li}_2(z)$ is the dilogarithm function.

Note that representations of $\phi_{2,0}(\alpha,\theta)$ differ from each other by the admixture of the hyperspherical harmonic (HH) $Y_{21}(\alpha,\theta)\propto \sin \alpha \cos \theta$ \cite{LEZ3}, where the hyperspherical angles $\alpha$ and $\theta$ are defined by Eq. (\ref{20})
and should not be confused with the $\alpha$ defined by Eq. (\ref{32}) for the \emph{collinear} configuration only.
The single-valued AFC $\tilde{\phi}_{2,0}(\alpha,\theta)$ (with no admixture of $Y_{21}$) was obtained, for $\phi_{2,0}(\alpha,\theta)$ defined by Eq. (22) of Ref. \cite{LEZ3}, in the form $\tilde{\phi}_{2,0}=\phi_{2,0}- \tilde{C}_{21}Z\sin \alpha \cos \theta$, where $\tilde{C}_{21}\simeq0.315837352$.

We would like to emphasize three points:

\begin{enumerate}
\item{} the AFC $\phi_{2,0}(\alpha,\theta)$ is not completely determined, as long as the contribution of the HH $Y_{21}(\alpha,\theta)$ admixture remains uncertain;

\item{} the contribution of that admixture, characterized by the coefficient $a_{21}$, was not calculated earlier;

\item{} below we shall propose a method of calculating $a_{21}$, at least for the ground states of the two-electron atomic systems.
\end{enumerate}

It was mentioned in Section \ref{S0} that we used two methods for the calculation of the WFs and the corresponding expectation values
of the two-electron atomic systems.
As the main method we applied the PLM \cite{LEZ1,LEZ2} with the number of shells $\Omega=25$.
In order to make sure that the obtained results are correct, we selectively used the CFHHM \cite{HM1,HM2} with the maximum  HH indices $K_m=128$ (1089 HH basis functions) for $\textrm{H}^-$, and $K_m=96$ (625 HH basis functions) for helium and the positive ions. In particular, the WFs with \emph{collinear} arrangement were calculated for the boundary cases of $\lambda=-1,0,1$ by the CFHHM.
For the basis sizes used, the first method generates the more accurate expectation values and energies. However, the near-the-origin behavior ($R\rightarrow 0$) behavior of the WFs calculated with the CFHHM is more accurate.
This can be explained by the fact that the radial parts of the WFs in the CFHHM are calculated
by numerical integration of exact regularized radial equations, while
the basis functions of the PLM do not include the logarithmic terms which are important for the correct representation of the WFs near the nucleus located at the origin.

Note that there is only one case of the \emph{collinear} arrangement where the AFC $\phi_{2,0}(\alpha,\theta)$ is independent of the admixture of $Y_{21}(\alpha,\theta)$. This is the case of the electron-nucleus coalescence where $\alpha=0$, and hence $\lambda=0$, and the theoretical expression (\ref{29}) becomes
\begin{equation}\label{34}
\xi_0=\frac{1-2E}{12}-Z\left(\frac{3-\ln 2}{6}\right)+\frac{1}{3}Z^2.
\end{equation}
For example, for the ground state of helium the expression (\ref{34}) yields $\xi_0\simeq 1.13167$.
The numerical CFHHM yields $\xi_0\simeq 1.13168$, whereas the PLM yields only $\xi_0\simeq 1.20558$.
For the two-electron ions under consideration we have obtained similar results.
This gives us substantial reasons to assume that the CFHHM generates the \emph{collinear} WFs with highly accurate behavior near the nucleus not only for the case $\lambda=0$, but for any value of $\lambda$, and for the boundary values of $\lambda=\pm1$ in particular (and may do so, in general, for any set of angles $\alpha$ and $\theta$).
For the latter cases the formula (\ref{29}) yields
\begin{equation}\label{35}
\xi_{\pm1}=2\left\{\frac{1-2E}{12}+Z\left[\omega^{(\pm)}\left(\frac{\pi}{2}\right)\pm a_{21} \right]
+\frac{5}{6}Z^2
\right\},
\end{equation}
where
\begin{equation}\label{36}
\omega^{(+)}\left(\frac{\pi}{2}\right)=\frac{1-\ln 2}{6},~~~~~~~~~
\omega^{(-)}\left(\frac{\pi}{2}\right)=\frac{3\ln 2-7}{6}.
\end{equation}
Note that the convergence of the Fock expansion was proved earlier \cite{MORG1}.
Using this convergence we can fit the truncated (up to $k=3$) Fock expansion of the form (\ref{25})
to the CFHHM collinear WFs for $\lambda=\pm1$  in the range $r\in [0,r_m]$  with $r_m\in [0.0001-0.005]$.
This enables us to obtain the numerical values of the coefficients $\xi_1$ and $\xi_{-1}$.
It is worth noting that only two parameters $\xi_{\pm 1}$ and $c_{\pm 1}^{(30)}$ were used as the fitted parameters, whereas the parameters $\eta_{\pm 1},~\zeta_{\pm 1}$ and $\gamma_{\pm 1}$ were calculated by Eqs. (\ref{26}), (\ref{27}) and (\ref{28}), respectively.
Substituting the obtained values of $\xi_{\pm1}$ into Eqs. (35), (36) we get two values of $a_{21}$ corresponding to $\xi_1$ and $\xi_{-1}$, respectively.
We would like to emphasize that these two values of $a_{21}$ coincide to within  4-5 significant digits, which confirms our initial assumption.
The coefficients $a_{21}$ so obtained are presented in Table \ref{T2}.
It is seen that these coefficients are different for every term of the helium-like isoelectronic sequence, and correspond to the ground states only.
One should emphasize that the obtained $a_{21}$ correspond to the specific form of the AFC $\phi_{2,0}(\alpha,\theta)$ represented by Eq. (22) from Ref. \cite{LEZ3}. However, using the admixture coefficient $\tilde{C}_{21}$ for the given representation of $\phi_{2,0}(\alpha,\theta)$, it is easy to calculate $a_{21}$ for any other representation, using the values presented in Table \ref{T2}.

\section{Analytic wave functions for the collinear arrangement} \label{S4}

The numerical calculations of individual terms of the Schr\"{o}dinger equation (\ref{1})
by the PLM \cite{LEZ1,LEZ2}
show that the collinear $S$-state WF,  $\Phi( |\lambda|r,r,(1-\lambda)r)$  can be represented with high accuracy by the solution $F_\lambda(r)$ of the differential equation:
\begin{equation}\label{37}
\left[\frac{d^2}{dr^2}+\left(\frac{A}{r}+B\right)\frac{d}{dr}+\left(\frac{C}{r}+D\right)\right]F_\lambda(r)=0,
\end{equation}
where the coefficients $A,~B,~C$ and $D$ are to be determined.
The general solution of Eq. (\ref{37}) is of the form:
\begin{equation}\label{38}
F_\lambda(r)=e^{-\frac{1}{2}r(B+\sigma)}\left[c_1 U(\kappa,A,\sigma r)+c_2 L_{-\kappa}^{(A-1)}(r \sigma)\right],
\end{equation}
where
\begin{equation}\label{39}
\sigma=\sqrt{B^2-4D},~~~~~~~~~\kappa=\frac{A}{2}+\frac{A B-2C}{2\sigma}.
\end{equation}
Here $U(\kappa,A,z)$ is the confluent hypergeometric function of the second kind (or the Tricomi function), and $L_{-\kappa}^{(A-1)}(z)$ is the generalized Laguerre function.

It can be verified that the series expansion of the generalized Laguerre function $L_{-\kappa}^{(A-1)}(z)$ at $z\rightarrow0$ does not contain terms with $\ln z$, whereas the Tricomi hypergeometric function $U(\kappa,A,z)$ contains logarithmic terms of the form $z^n \ln z$ with integer $n$, but only if the parameter $A$ is also an integer.
And lastly, only the series expansion of the Tricomi hypergeometric function with parameter $A=-1$  contains the terms proportional to $z^n \ln z$ with $n\geq 2$ similar to the Fock expansion (\ref{19}).
There is only one exception, occurring when the leading term of the logarithmic series of the Fock expansion is proportional to $R^3\ln R$. This happens at the electron-nucleus coalescence (when the \emph{collinear} parameter $\lambda=0$), and will be considered separately.

The general solution of Eq. (\ref{37}) for $A=-1$ is of the specific form:
\begin{equation}\label{40}
F_\lambda(r)=e^{-\frac{1}{2}r(B+\sigma)}\left[c_1 U(\kappa,-1,\sigma r)+c_2 r^2 L_{-\kappa-2}^{(2)}(\sigma r)\right].
\end{equation}
Note that
\begin{equation}\label{41}
U(\kappa,-1,0)=1/\Gamma(\kappa+2),~~~~~~~ L_{-\kappa-2}^{(2)}(0)=\kappa(\kappa+1)/2,
\end{equation}
whereas the asymptotic behavior of the special functions on the RHS of the solution (\ref{40}) is defined by the following series expansions:
\begin{equation}\label{42}
U(\kappa,-1,\sigma r)\underset{r\rightarrow \infty}{=}(\sigma r)^{-\kappa}
\left[1-\frac{\kappa(\kappa+2)}{\sigma r}+
\frac{\kappa(\kappa+1)(\kappa+2)(\kappa+3)}{2(\sigma r)^2}+O(r^{-3})\right],
\end{equation}
\begin{eqnarray} \label{43}
L_{-\kappa-2}^{(2)}( \sigma r)\underset{r\rightarrow \infty}{=}
r^{-\kappa}\left[\frac{\kappa(\kappa+1)(-\sigma)^{-\kappa-2}}{\Gamma(1-\kappa)r^2}+O(r^{-3})\right]+
~\nonumber~~~~~~~~~~~~~~~~~~~~~~~~~~~~~~~~\\
r^\kappa e^{\sigma r}\frac{(\kappa+1)\kappa\sigma^{\kappa-2}}{\Gamma(\kappa+2)}\left[\frac{\sigma}{r}+\frac{\kappa^2-1}{r^2}+O(r^{-3})\right].
\end{eqnarray}
The second asymptotic expansion implies that the function (\ref{40}) with $c_2\neq 0$ can possess the physical property of exponential decay at $r\rightarrow \infty$ only if $B>\textrm{Re}(\sigma)$
with real parameter $B$.
Moreover, it follows from Eqs. (\ref{40}), (\ref{41}) that we cannot set $c_1=0$ in Eq. (\ref{40})
for $F_\lambda(r)$ representing the $S$-state WF.
For the case under consideration there are only two possibilities: either to set $c_2=0$ or to set $c_1\neq 0$ and $c_2\neq 0$.
First we shall consider the simpler case $c_2=0$.
For this case the analytic WF (\ref{40}), satisfying the condition $F_\lambda(0)=1$ (like the Fock expansion (\ref{19})), becomes
\begin{equation}\label{44}
F_\lambda(r)=\Gamma(\kappa+2)e^{-\frac{1}{2}r(B+\sigma)} U(\kappa,-1,\sigma r),~~~~~~~~~~~~~(0<|\lambda|\leq1).
\end{equation}
This model WF contains \emph{three} unknown parameters $B,~C$ and $D$.
In order to calculate these parameters we need to obtain three coupling equations for them.
We propose to derive 2 of the 3 required  equations
by employing the Fock expansion (\ref{19}) describing the behavior of the WF in the vicinity of the nucleus ($r\rightarrow 0$).

The series expansion of the function (\ref{44}) near the origin is of the form
\begin{equation}\label{45}
F_\lambda(r)\underset{r\rightarrow 0}{=}1-\frac{1}{2}\left[B+(2\kappa+1)\sigma\right] r-\frac{1}{2}\kappa(\kappa+1)\sigma^2r^2 \ln r+C_2r^2+O(r^3),
\end{equation}
where the coefficient $C_2$ is rather complicated and, what is more important, we will not apply it in our upcoming consideration.

It was shown in Section \ref{S3} that the Fock expansion of the WF with \emph{collinear} configuration is defined by Eq. (\ref{25}), where the expansion coefficients $\eta_\lambda,~\zeta_\lambda,~\gamma_\lambda$ and $\xi_\lambda$  can be calculated by Eqs. (\ref{26}), (\ref{27}), (\ref{28}) and (\ref{29}), respectively.
Equating coefficients for $r$ and $r^2\ln r$ in expansions (\ref{45}) and (\ref{25}), we obtain the first two coupling equations we were looking for:
\begin{equation}\label{46}
B+(2\kappa+1)\sigma + 2\eta_\lambda=0,~~~~~~\kappa(\kappa+1)\sigma^2+2\zeta_\lambda=0.
\end{equation}
Eq. (\ref{46}) enables us to express the parameters $\sigma$ and $\kappa$ in terms of parameter $B$.
Denoting for simplicity $\eta_\lambda=\eta$, and $\zeta_\lambda=\zeta$, we obtain:
\begin{equation}\label{47}
\sigma=\varrho,~~~~~\kappa=-\frac{1}{2}-\frac{B+2\eta}{2\varrho},
\end{equation}
where
\begin{equation}\label{48}
\varrho=\sqrt{8\zeta+(B+2\eta)^2}.
\end{equation}
Note that Eq. (\ref{47}) represents the solution of the set of Eqs. (\ref{46}) for $\textrm{Re}(\sigma)>0$ according to the definition (\ref{39}).

Inserting the parameters $\sigma$ and $\kappa$ defined by Eqs. (\ref{47}) into  Eq. (\ref{44}), we obtain the model WF as a function of a single parameter $B$ which can be calculated as follows.
Remember that the first two coupling equations were derived by the use of the well-known behavior of the WF near the origin ($r\rightarrow 0$).
The natural and also the correct way to find the third (and efficient) coupling equation is to use the integral and/or asymptotic ($r\rightarrow \infty$) properties of the actual (numerically calculated three-body) WF.
We shall return later to the application of some asymptotic properties required in the consideration of the more complicated WF. As to the model WF of the form (\ref{44}), we then propose to employ the expectation value $\left\langle\delta\left(\textbf{r}_1-\lambda \textbf{r}_2\right)\right\rangle$ as the third coupling equation  enabling us to find the parameter $B$. Remember that according to Eqs. (\ref{11}) and (\ref{7}) the value of $\left\langle\delta\left(\textbf{r}_1-\lambda \textbf{r}_2\right)\right\rangle$ is defined by the collinear WF, $\tilde{\Phi}(|\lambda|r,r,(1-\lambda)r)$ and the expectation value $\langle\delta(\textbf{r}_1)\delta(\textbf{r}_2)\rangle$
evaluated in the three-body space.
Consequently, the third coupling equation has the form
\begin{equation}\label{49}
\left\langle\delta\left(\textbf{r}_1-\lambda \textbf{r}_2\right)\right\rangle=
4\pi \langle\delta(\textbf{r}_1)\delta(\textbf{r}_2)\rangle\int_0^\infty \left|F_\lambda (r)\right|^2 r^2 dr.
\end{equation}
Note that in some cases the function (\ref{44}) can be complex, therefore the integrand includes the square of its absolute value.

The values of $\langle\delta(\textbf{r}_1)\delta(\textbf{r}_2)\rangle$ which represent, in fact, the values of the square of the normalized WF at the nucleus, can be found in Refs. \cite{FR1,FR2,FR3} (see also references therein).
The values of $\left\langle\delta\left(\textbf{r}_1-\lambda \textbf{r}_2\right)\right\rangle$ are presented in a number of publications (see, e.g., \cite{DRK,FR1,FR2,FR3} and references therein), but only for $\lambda=0$ and $\lambda=1$, i.e., for the two-particle atomic coalescences.
In Table \ref{T1} we present the expectation values $\left\langle\delta\left(\textbf{r}_1-\lambda \textbf{r}_2\right)\right\rangle$ for $\lambda=-1,-0.5,0,0.5,1$ calculated by the PLM (with $\Omega=25$).

The model WF (\ref{44}) provides a highly accurate approximation of the actual WF for almost all of the considered atomic systems, at least for $\lambda=\pm 1, \pm 1/2$.
The corresponding parameters $A,~B,~C$ and $D$, as well as the auxiliary parameters $\sigma$ and $\kappa$, calculated by the method mentioned above, are presented in Table \ref{T3} for helium and all of the considered ions.

Exceptions that have to be considered separately are the negative $\textrm{H}^-$ ion for $\lambda=\pm 1/2$, and the case of $\lambda=0$ in general.
To obtain quite an accurate model WF for the $\textrm{H}^-$ ion with collinear configuration corresponding to the collinear parameter $\lambda=\pm 1/2$, it suffices to extend expression (\ref{44}) by the Laguerre function
 $L_{-\kappa}^{(m)}$ with a specific integer parameter $m$. Thus, for $Z=1$ only, we obtain
\begin{equation}\label{50}
F_{\frac{1}{2}}(r)=e^{-\frac{1}{2}r(B+\sigma)}\left[\Gamma(\kappa+2) U(\kappa,-1,\sigma r)+c_+ L_{-\kappa}^{(-1)}(\sigma r)\right],
\end{equation}
\begin{equation}\label{51}
F_{-\frac{1}{2}}(r)=(1+c_2)^{-1}e^{-\frac{1}{2}r(B+\sigma)}\left[\Gamma(\kappa+2) U(\kappa,-1,\sigma r)+
c_-\left(\frac{720 \Gamma(1-\kappa)}{\Gamma(7-\kappa)}\right) L_{-\kappa}^{(6)}(\sigma r)\right].
\end{equation}
Equating the coefficients for $r$ and $r^2\ln r$ in the Fock expansion (\ref{25}) and the series expansion of (\ref{50}), we obtain two coupling equations of the form
\begin{equation}\label{52}
B+(1+2\kappa+2 c_{+})\sigma=-2\eta,~~~~~~~~~\kappa(1+\kappa)\sigma^2=-2\zeta,
\end{equation}
where $\eta=-5/4$ and $\zeta=(2-\pi)/(3\pi)$ for $Z=1$ and $\lambda=1/2 $ (see Eqs. (\ref{26}) and (\ref{27})).

The similar procedure for the function (\ref{51}) yields another two coupling equations
\begin{equation}\label{53}
\sigma\left[\frac{2 \kappa (7-c_-)}{7(1+c_{-})}+1\right]+B=-2\eta,
~~~~~~\kappa(1+\kappa)\sigma^2=-2(1+c_-)\zeta,~~
\end{equation}
where $\eta=-3/4$ and $\zeta=(\pi-2)/(3\pi)$ for  $Z=1$ and $\lambda=-1/2 $.

As was mentioned earlier an extra coupling equation should be derived from the asymptotic behavior of the physical WF.
Accordingly, we would like to recall that many authors (see, e.g., \cite{FOCK}, \cite{PEK1}, \cite{AHL2}, \cite{AHL}) found it correct and efficient to look for the singlet physical WFs in the form
\begin{equation}\label{54}
\Psi= e^{(-a r_1-b r_2)}G(r_1,r_2,r_{12})+e^{(-a r_2-b r_1)}G(r_2,r_1,r_{12}),
\end{equation}
where
\begin{equation}\label{55}
a=\sqrt{-2E-Z^2},~~~~~~~~~~b=Z.
\end{equation}
Representation (\ref{54})-(\ref{55}) along with the asymptotic expansion of the Laguerre function $L_{-\kappa}^{(m)}(\sigma r)\propto \exp(\sigma r)$ enables us to obtain the following extra coupling equation for the model WFs (\ref{50}) and (\ref{51}),
\begin{equation}\label{56}
B-\sigma=2 \min(|\lambda|a+b,|\lambda|b+a),
\end{equation}
where the parameters $a$ and $b$ are defined by Eq. (\ref{55}).

Thus, we have obtained three coupling equations, given by Eqs. (\ref{52}), (\ref{53}) and (\ref{56}), for the four parameters $B,~\sigma,~\kappa$ and $c_{\pm}$  corresponding to $Z=1$ and the \emph{collinear} parameter $\lambda=\pm 1/2$. To calculate these parameters we need one more coupling equation which can be represented, as previously, by the integral relation (\ref{49}).
The numerical parameters describing the collinear WFs (\ref{50}) and (\ref{51}) are presented in Table \ref{T4}.

The last point we would like to discuss is the specific case of the electron-nucleus coalescence corresponding to the \emph{collinear} parameter $\lambda=0$.
As noted previously, in this case the leading term of the logarithmic series of the Fock expansion is
proportional to $R^3\ln R$. Hence, to describe correctly the near-the-origin behavior of the model WF, $F_0(r)$ we should use the Tricomi function $U(\kappa,A,\sigma r)$ with $A=-2$ (instead of $A=-1$ used previously).
In general, it was stated that the most accurate approximation can be obtained by the model function of the form
\begin{equation}\label{57}
F_0(r)=(1+c_0)^{-1}e^{-\frac{1}{2}r(B+\sigma)}
\left[\frac{1}{1-\kappa} L_{-\kappa}^{(1)}(\sigma r)+c_0\frac{\Gamma(\kappa+3)}{2} U(\kappa,-2,\sigma r) \right].
\end{equation}
with the four parameters $B,~\sigma,~\kappa,~c_0$ currently undetermined. Note that in this case the two leading terms of the \emph{collinear} Fock expansion (\ref{25}) become the terms proportional to $r$ and $r^2$, because the term $\zeta_0r^2\ln r$ disappears due to $\zeta_0=0$.
Thus, equating the coefficients for $r$ and $r^2$ in the expansion (\ref{25}) and the series expansion for the function (\ref{57}), we obtain two coupling equations of the form
\begin{equation}\label{58}
B+\sigma\left[1+\frac{\kappa(c_0-1)}{c_0+1}\right]=-2\eta_0,
\end{equation}
\begin{equation}\label{59}
B\left[B+2\sigma\left(1+\frac{\kappa(c_0-1)}{c_0+1}\right)\right]+
\sigma^2\left[1+\frac{2\kappa[\kappa-2+3c_0(\kappa+2)]}{3(c_0+1)}\right]=8\xi_0,
\end{equation}
where $\eta_0=1/2-Z$, and $\xi_0$ is defined by Eq. (\ref{34}).
The asymptotic relation (\ref{56}), which (for $\lambda=0$) reduces to $B-\sigma=2a$, can be used as the third coupling equation. And at last, as previously, the integral relation (\ref{49}) can be employed as the fourth coupling equation needed for calculation of the parameters $B,~\sigma,~\kappa$ and $c_0$ of the model WF (\ref{57}).
These parameters are presented in Table \ref{T4} for all members of the helium-like isoelectron sequence considered in this paper.

\begin{figure}
\caption{The ground state of the negative $\textrm{H}^-$ ion: \textbf{(\emph{a})} the WF at the electron-nucleus coalescence (the \emph{collinear} parameter $\lambda=0$), times $r$; \textbf{(\emph{b})} the logarithmic estimate of the difference between the model WF (\ref{57}) and the PLM WF represented by the solid line (\red{red} online), and between the CFHHM WF and the PLM WF represented by the dashed line (\blue{blue} online).}
\includegraphics[width=6.0in]{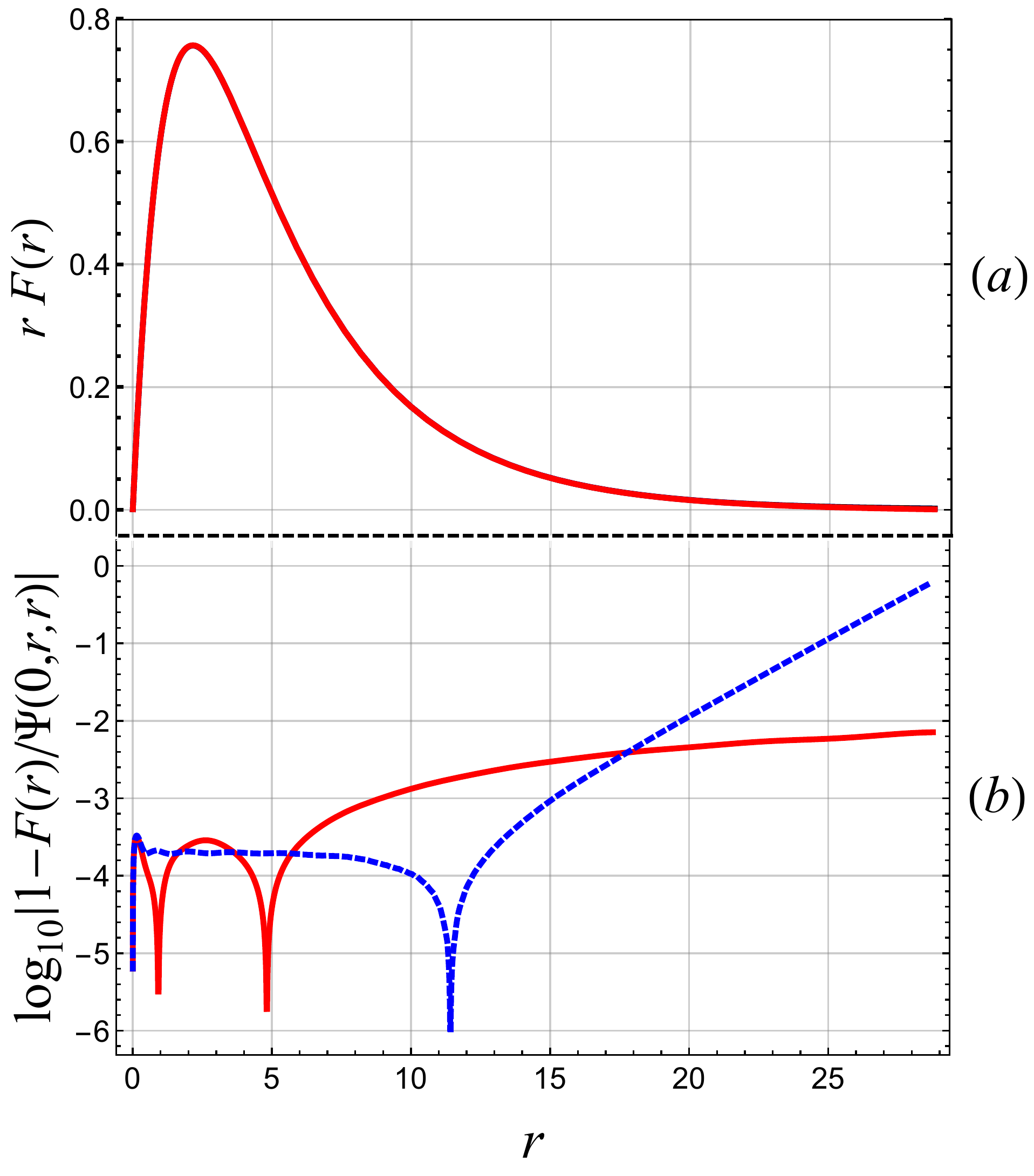}
\label{F3}
\end{figure}

To estimate the difference between two functions $\mathcal{F}(r)$ and $\mathcal{P}(r)$ or the accuracy of one of them, we used the logarithmic expression
\begin{equation}\label{60}
\mathcal{L}=\log_{10}\left|1-\mathcal{F}(r)/\mathcal{P}(r)\right|.
\end{equation}
In Fig. \ref{F3}(\emph{a}) we display three functions (times $r$) representing the ground state of the negative $\textrm{H}^-$ ion at the electron-nucleus coalescence when the collinear parameter  $\lambda=0$. These functions are: the PLM WF, $\tilde{\Phi}_{PLM}(0, r,r)$, the CFHHM WF,  $\tilde{\Phi}_{CFHHM}(0, r,r)$, and the model WF, $F_0(r)$ defined by Eq. (\ref{57}). It is seen that there is no visible difference between either pair of functions because of their extreme proximity which is demonstrated by  Fig. \ref{F3}(\emph{b}). In the latter figure we show two functions of the form (\ref{60}).
The first function $\mathcal{L}_1$ represented by solid line (red online) describes the accuracy of the model WF
(\ref{57}), that is $\mathcal{F}(r)=F_0(r)$ and $\mathcal{P}(r)=\tilde{\Phi}_{PLM}(0,r,r)$, whereas $\mathcal{L}_2$ represented by dashed line (blue online) describes the difference between the WFs calculated with the CFHHM and the PLM, that is  $\mathcal{F}(r)=\tilde{\Phi}_{CFHHM}(0,r,r)$ and $\mathcal{P}(r)=\tilde{\Phi}_{PLM}(0,r,r)$.
It is seen that for $0<r<6$ and for $r>17.5$ (in a.u.) the curve $\mathcal{L}_1$ lies even lower than $\mathcal{L}_2$, which demonstrates the extremely high accuracy of the model WF (\ref{57}), at least for the $\textrm{H}^-$ ion.

\begin{figure}
\caption{The ground state of the helium atom and the positive ions with $Z=3,4,5$: \textbf{(\emph{a})}
the PLM WFs, $\mathbf{\Phi}_0^{(Z)}(\rho)\equiv\tilde{\Phi}(0,\rho/Z,\rho/Z)$, at the electron-nucleus coalescence (the \emph{collinear} parameter $\lambda=0$) times $\rho/Z $; \textbf{(\emph{b})} the logarithmic estimate of the difference between the model WFs, $F_0(\rho/Z)$, and the PLM WFs.}
\includegraphics[width=6.0in]{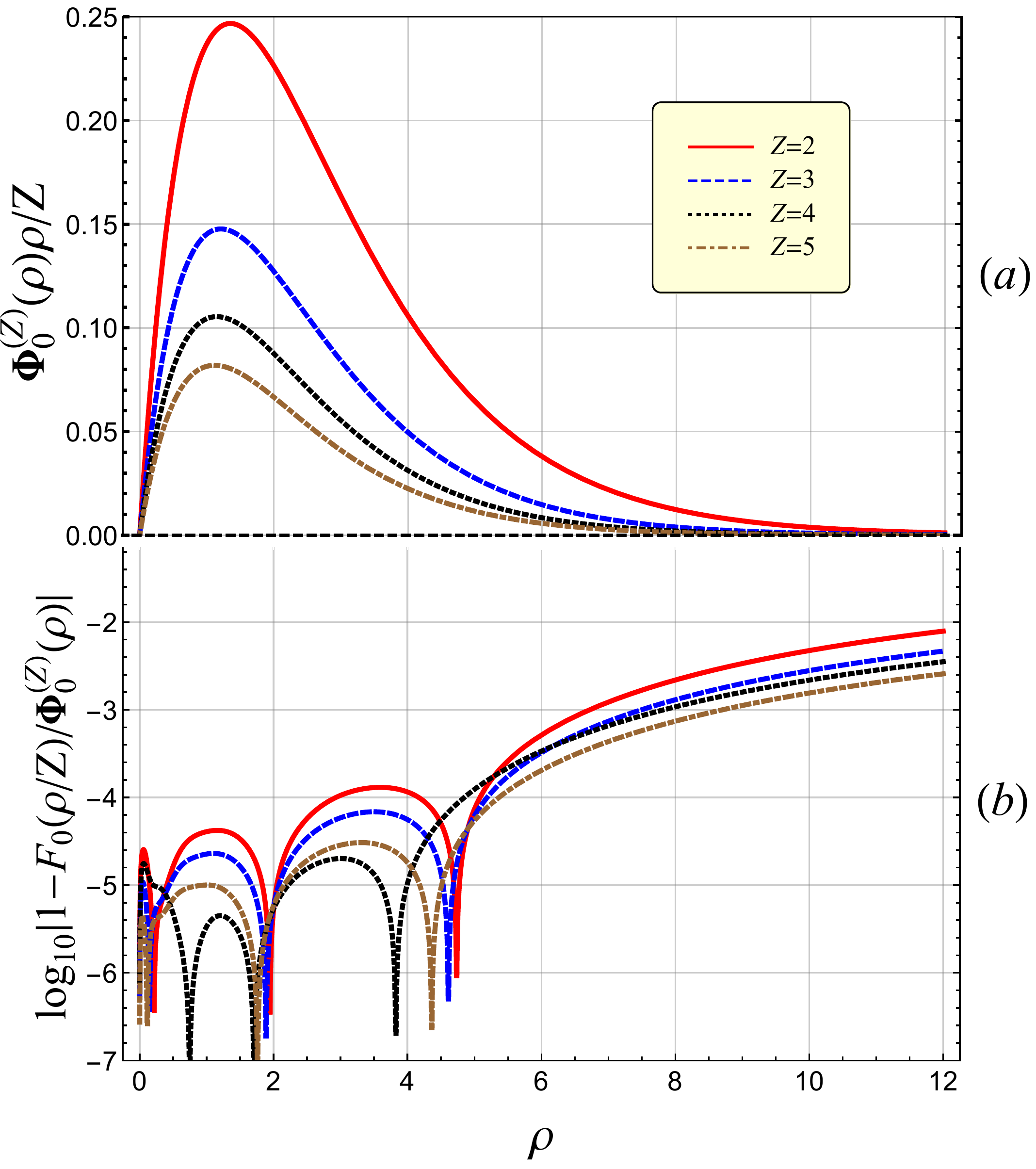}
\label{F4}
\end{figure}

In Fig. \ref{F4}(\emph{a}) we display the PLM WFs, $\mathbf{\Phi}_0^{(Z)}(\rho)\equiv\tilde{\Phi}(0,\rho/Z,\rho/Z)$ at the electron-nucleus coalescence ($\lambda=0$) times $\rho/Z$ for the helium atom and positive (two-electron) ions with
$Z=3,4,5$.
In this and the following figures, to accommodate all functions on the same scale, we plot $\rho=Z r$ instead of $r$ on the abscissa.
In Fig. \ref{F4}(\emph{b}) the logarithmic differences (\ref{60}) between the model WFs, $F_0(\rho/Z)$ and the PLM WFs, $\mathbf{\Phi}_0^{(Z)}(\rho)$ are shown, demonstrating the high accuracy of the model WFs of the form (\ref{57}) constructed for the specific \emph{collinear} case of $\lambda=0$.

\begin{figure}
\caption{The ground state of the helium-like systems with $1\leq Z\leq 5$: \textbf{(\emph{a})}
the PLM WFs, $\mathbf{\Phi}_1^{(Z)}(\rho)\equiv\tilde{\Phi}(\rho/Z,\rho/Z,0)$, at the electron-electron coalescence (the \emph{collinear} parameter $\lambda=1$) times $\rho/Z $; \textbf{(\emph{b})} the logarithmic estimate of the difference between the model WFs, $F_1(\rho/Z)$, and the PLM WFs.}
\includegraphics[width=6.0in]{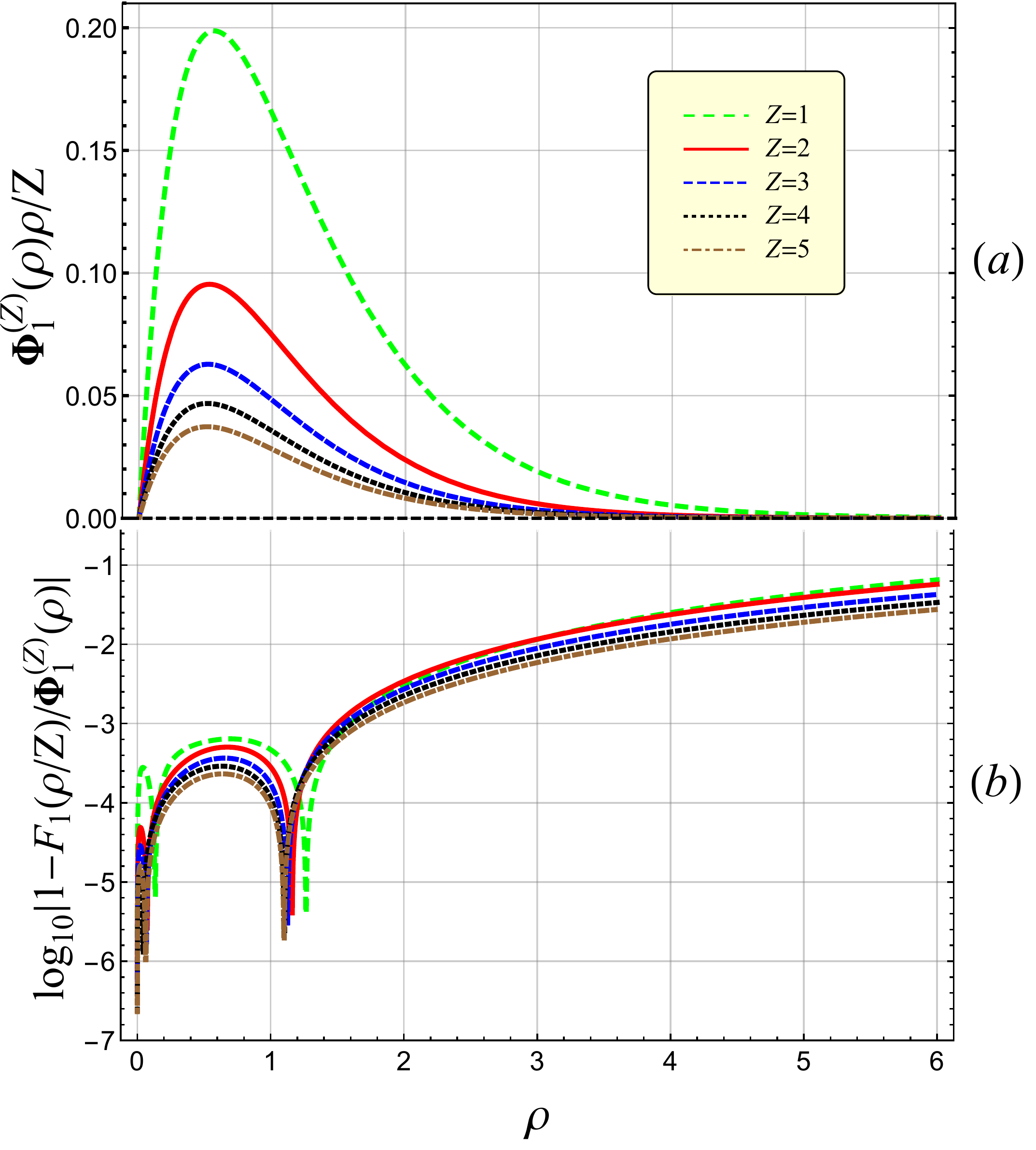}
\label{F5}
\end{figure}

In Fig. \ref{F5}(\emph{a}) we display the PLM WFs, $\mathbf{\Phi}_1^{(Z)}(\rho)\equiv\tilde{\Phi}(\rho/Z,\rho/Z,0)$ at the electron-electron coalescence ($\lambda=1$) times $\rho/Z$ for the helium-like atoms with $1\leq Z\leq5$.
 In Fig. \ref{F5}(\emph{b}) the logarithmic differences (\ref{60}) between the model WFs, $F_1(\rho/Z)$ and the PLM WFs, $\mathbf{\Phi}_1^{(Z)}(\rho)$ are shown, demonstrating the high accuracy of the model WFs of the form (\ref{44}) as applied to the specific \emph{collinear} case of $\lambda=1$.

\begin{figure}
\caption{The ground state of the helium-like systems with $1\leq Z\leq 5$: \textbf{(\emph{a})}
the PLM WFs, $\mathbf{\Phi}_{-1}^{(Z)}(\rho)\equiv\tilde{\Phi}(\rho/Z,\rho/Z,2\rho/Z)$, at the \textbf{e-n-e} \emph{collinear} configuration ($\lambda=-1$) times $\rho/Z $; \textbf{(\emph{b})} the logarithmic estimate of the difference between the model WFs, $F_{-1}(\rho/Z)$, and the PLM WFs.}
\includegraphics[width=6.0in]{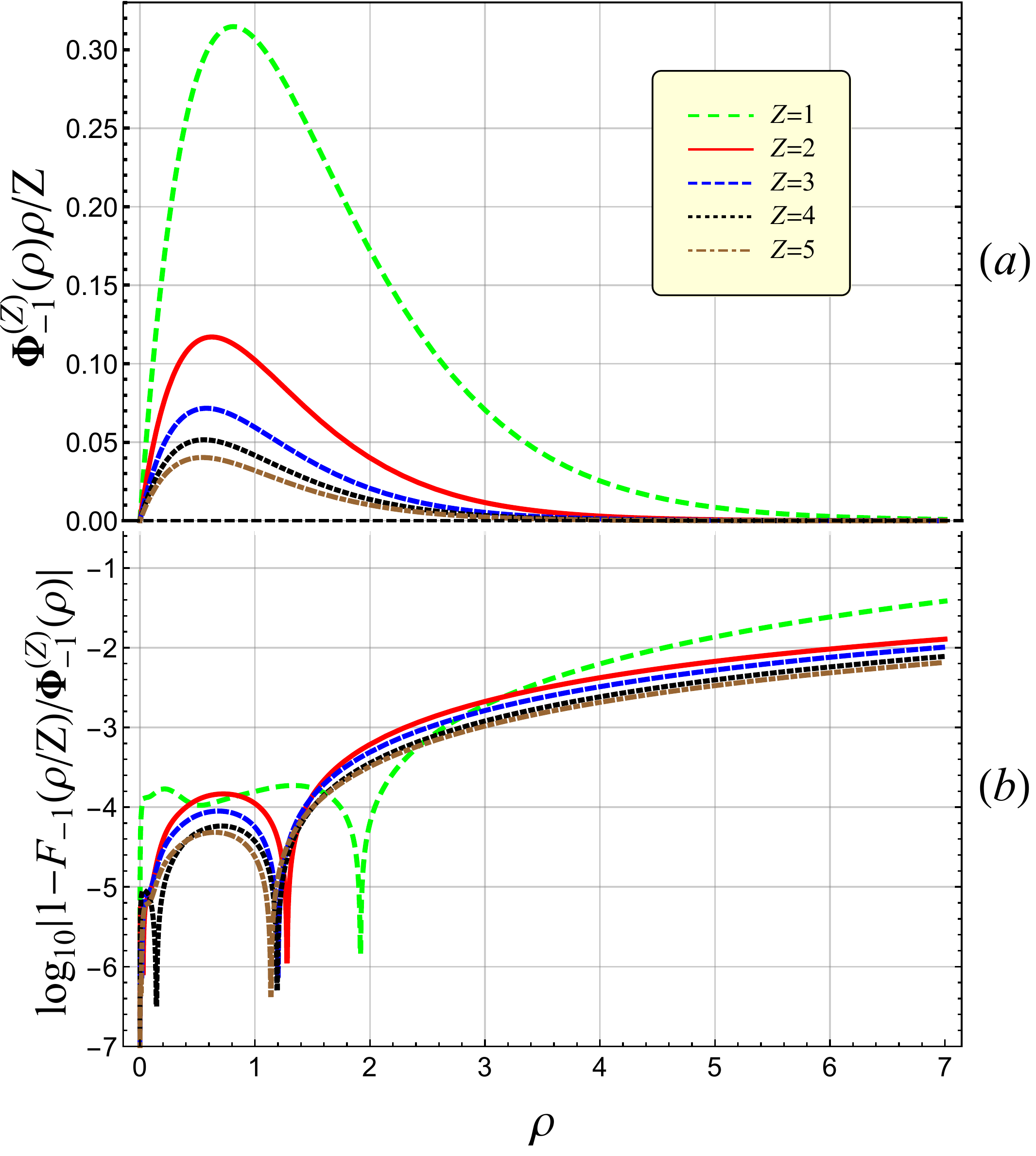}
\label{F6}
\end{figure}

In Fig. \ref{F6}(\emph{a}) we display the PLM WFs, $\mathbf{\Phi}_{-1}^{(Z)}(\rho)\equiv\tilde{\Phi}(\rho/Z,\rho/Z,2\rho/Z)$ at the \textbf{e-n-e} \emph{collinear} configuration ($\lambda=-1$) times $\rho/Z$ for the helium-like atoms with $1\leq Z\leq5$.
In Fig. \ref{F6}(\emph{b}) the logarithmic differences (\ref{60}) between the model WFs, $F_{-1}(\rho/Z)$ and the PLM WFs, $\mathbf{\Phi}_{-1}^{(Z)}(\rho)$ are shown, demonstrating again the high accuracy of the model WFs of the form (\ref{44}) as applied to the specific \emph{collinear} case of $\lambda=-1$.

\section{Conclusions} \label{S5}

We have investigated the properties of the helium-like isoelectron sequence with the \emph{collinear} arrangement of particles.
The two-electron atomic systems with nucleus charge $1\leq Z\leq5$, i.e., the negative $\textrm{H}^-$ ion, the helium atom and the positive ions $\textrm{Li}^+,~\textrm{Be}^{2+}$ and $\textrm{B}^{3+}$ were taken as examples.
Two fully three-body methods, the PLM \cite{LEZ1, LEZ2} and the CFHHM \cite{HM1, HM2}, were used to calculate the ground states and the corresponding expectation values.

The collinear configurations were parameterized by the single scalar parameter $-1\leq \lambda\leq1$.
The particular case $\lambda=0$ corresponds to the electron-nucleus ($\textbf{e-n}$)
coalescence, whereas $\lambda=1$ corresponds to the electron-electron ($\textbf{e-e}$) coalescence.
In general, $\lambda>0$ corresponds to the \textbf{n-e-e} configuration, and $\lambda<0$ to the \textbf{e-n-e} configuration.
It was derived that, at least for the $S$-state, the expectation value  $h(\lambda,Z)\equiv\left\langle\delta\left(\textbf{r}_1-\lambda \textbf{r}_2\right)\right\rangle$  depends only on the collinear wave function $\tilde{\Phi}(|\lambda|r,r,(1-\lambda)r)$ and the expectation value $\langle\delta(\textbf{r}_1)\delta(\textbf{r}_2)\rangle$ which can be found in Refs.\cite{FR1, FR2, FR3}. The specific cases of $\left\langle\delta\left(\textbf{r}_1-\lambda \textbf{r}_2\right)\right\rangle$ corresponding to the electron-nucleus ($\lambda=0$) and electron-electron ($\lambda=1$) coalescences have been calculated and published earlier in a number of articles (see, e.g., \cite{DRK, FR1, FR2, FR3} and references therein).
The value of $\left\langle\delta\left(\textbf{r}_1+\textbf{r}_2\right)\right\rangle$ corresponding to the other boundary  $\lambda=-1$,
as well as the values of $\left\langle\delta\left(\textbf{r}_1-\lambda\textbf{r}_2\right)\right\rangle$ for some other $\lambda$, have been
calculated and presented for the first time in the current work along with some specific values of $\lambda$ (see Table \ref{T1}).
The general dependence of the $\lambda$-normalized expectation values
$\Gamma_Z(\lambda)\equiv h(\lambda,Z)/M(Z)$ (see also Eq.(\ref{11a})) on the \emph{collinear} parameter $\lambda\in[-1,1]$ is presented in Fig. \ref{F1} for all the atomic systems under consideration.
Simple analytic expression (\ref{11b}) for the asymptotic curve $\Gamma_\infty(\lambda)$ has been derived and presented in Fig. \ref{F1}, as well.

Simple formulas for calculating the expectation values of the kinetic and potential energy operators $\left\langle \delta T\right\rangle$ and $\left\langle \delta V\right\rangle$ for the $S$-state with the collinear arrangement of the particles were derived, and the results were presented in Table \ref{T2} and Fig. \ref{F2}.
These expectation values exhibit a few characteristic points located at $\lambda$ values listed in Sec. \ref{S2}.
Unusual physical properties of the \textbf{n-e-e} collinear configurations were found for certain ranges of the parameter $\lambda$.
In particular it was found that for the helium atom as well as for all the two-electron ions considered there exist specific $\lambda $ intervals where both $\left\langle \delta T\right\rangle$ and $\left\langle \delta V\right\rangle$ are negative. We have reasons to presume an important physical significance of these $\lambda$ intervals.

Analytic formulas for the first few angular Fock coefficients (AFCs) were obtained for the \emph{collinear} configurations parameterized by $\lambda$. Numerical values of the parameter $a_{21}$, completing the full description of the AFC $\phi_{2,0}$
in the part where analytic expressions are not available,
were calculated for helium atom and for each two-electron ion under consideration (see Sec. \ref{S3} and Table \ref{T2}).

Highly accurate model wave functions describing the ground states of the two-electron atoms with collinear arrangement were obtained. These model WFs
are expressed in terms of the Tricomi function (the confluent hypergeometric function of the second kind) and the generalized Laguerre functions, and are parameterized with 3 or 4 parameters. These
parameters are determined from (i) the analytic structure of the three-body wave function
near the origin and at infinity, and (ii) from the expectation values calculated from the
numerical three-body wave functions.


\section{Acknowledgment}

This work was supported by the PAZY Foundation.
We acknowledge helpful discussions with Prof. M. Ya. Amusia.

\newpage

\begin{table}
\caption{Expectation values $\left\langle\delta\left(\textbf{r}_1-\lambda \textbf{r}_2\right)\right\rangle$ for the two-electron atomic systems (with nuclear charge $Z$) in the ground state.
The ground state WFs were calculated by the PLM \cite{LEZ1,LEZ2} with the number of shells $\Omega=25$.
The corresponding energies $E$ are presented in the last line.
The results of the more accurate calculations are available \cite{DRK,FR1,FR2,FR3}. They can be obtained by replacing our last digit by the adjacent digit in square brackets.}
\begin{tabular}{|c|c|c|c|c|c|}
\hline
$\lambda \backslash Z$ & $1$ & $2$ & $3$ & $4$ & $5$\tabularnewline
\hline
\hline
-1 &0.009126167  &0.1827953& 0.7571887  &1.9716347 &4.0649088 \tabularnewline
\hline
-0.5 &0.02336386  &0.4437669& 1.8202429  &4.7202462 &9.7097718 \tabularnewline
\hline
0 & 0.1645527[9] &1.810429317[8] &  6.852008[9] &17.1981724[5]& 34.758743[4] \tabularnewline
\hline
0.5 & 0.01206460 &0.3293527 &  1.5016168 &4.0952366& 8.6761318 \tabularnewline
\hline
1 &0.00273806[0] &0.1063455[4] & 0.5337227[5] & 1.5228957[4] & 3.312443[1] \tabularnewline
\hline
\hline
$\lambda_{v2}$ & 0.018649 & 0.34767 & 1.4281 & 3.6692 & 7.4792 \tabularnewline
\hline
$\lambda_2$ & 0.01210173 & 0.2141994 & 0.8548817 & 2.175692 & 4.415435\tabularnewline
\hline
$\lambda_{t0}$ & 0.009520788 & 0.1837904 & 0.7597262 & 1.9766583 & 4.073548 \tabularnewline
\hline
$\lambda_{cr}$ & 0.008659116& 0.1739671& 0.7302144&1.915920& 3.970376 \tabularnewline
\hline
$\lambda_{v0}$ & 0.007979059 & 0.1664143 & 0.7074943 & 1.869972 & 3.892486 \tabularnewline
\hline
\hline
$M(Z)$ &0.065991408 & 1.0994792 & 4.5084853 &11.728245  & 24.191315\tabularnewline
\hline
$-E$&0.52775101652[4] &2.903724377033[4]  &7.279913412668[9]  &  13.6555662384231[5] &  22.0309715802421[8] \tabularnewline
\hline
\end{tabular}
\label{T1}
\end{table}

\begin{table}
\caption{The characteristic points of the
expectation values $\left\langle \delta V\right\rangle$ and $\left\langle
\delta T\right\rangle$ for the ground state of the two-electron atomic systems
with nuclear charge $Z$ as functions of the \emph{collinear} parameter $\lambda$.
The column before the last lists the ratios $\left\langle\delta V\right\rangle/\left\langle \delta T\right\rangle$ for $\lambda=-1$, and the last column gives the parameter $a_{21}$ of the AFC $\phi_{2,0}$ (see Eq. (\ref{29})).}
\begin{tabular}{|c|c|c|c|c|c|c|c|}
\hline
$Z\setminus\
$ & $\lambda_{v0}$ & $\lambda_{t0}$ & $\lambda_{2}$ & $\lambda_{cr}$ & $\lambda_{v2}$ & $\left\langle\delta V\right\rangle/\left\langle \delta T\right\rangle(\lambda=-1)$&$a_{21}$\tabularnewline
\hline
\hline
1 & 0.6181 & 0.5659 &  0.4992  & 0.5936 & 0.3896 & -1.3941 & -0.09028\tabularnewline
\hline
2 & 0.7811 &  0.7362  & 0.6701 &  0.7607  & 0.4801 &-1.3473 &-0.11915  \tabularnewline
\hline
3 & 0.8472 & 0.8106 & 0.7520 & 0.8308 & 0.5207&-1.3406 & -0.14784 \tabularnewline
\hline
4 &0.8828 &0.8524 & 0.8012 &0.8694 &0.5481 & -1.3381 & -0.1757\tabularnewline
\hline
5 & 0.9050  & 0.8791 & 0.8342 & 0.8937 & 0.5674 & -1.3369 & -0.2287\tabularnewline
\hline
\end{tabular}
\label{T2}
\end{table}

\begin{table}
\caption{The coefficients for the model WF of the form (\ref{44}) as functions of $Z$ and $\lambda$.}
\begin{tabular}{|c|c|c|c|c|c|c|c|}
\hline
$\lambda$ & $Z$ & $A$ & $B$ & $C$ & $D$ & $\sigma $& $\kappa$ \tabularnewline
\hline
\hline
-1 & 1 & -1 &  1.020138426  & -1 &-0.4643685438 & 1.702397305 & -0.2122112533\tabularnewline
 &2 &    & 3.589566964 &  -3  & 0.7996869527 &3.112272993 &-0.1127535982  \tabularnewline
 & 3 & &6.142518812 &-5 & 4.259073148 & 4.549092740 &-0.0760164270 \tabularnewline
 &4 && 8.698936064 &-7 &9.954524567 &5.987770067 &-0.0573423932\tabularnewline
 &5  && 11.24789303 &-9& 17.80850242 &  7.435125281 &-0.0459318629\tabularnewline
 \hline
-0.5 & 2 &-1  & 2.546341682  & -2.25 & 0.1822618141& 2.398918236 & -0.09280431311 \tabularnewline
     & 3 &    & 4.399223532  & -3.75 & 1.707827789 & 3.538623536 & -0.06186686194 \tabularnewline
     & 4 &    & 6.252397591  & -5.25 & 4.293573411 & 4.681685806 & -0.04635973183 \tabularnewline
     & 5 &    & 8.097830582  & -6.75 & 7.886589004 & 5.833395591 & -0.03696184893 \tabularnewline
 \hline
0.5  & 2 &-1  & 3.776185440  & -2.75 & 3.306516932 & 1.016616326 & 0.3478196321 \tabularnewline
     & 3 &    & 6.008971043  & -4.25 & 8.202387390 & 1.816090152 & 0.1858219437 \tabularnewline
     & 4 &    & 8.239464542  & -5.75 & 15.28343506 & 2.599045154 & 0.1272564086 \tabularnewline
     & 5 &    & 10.47361342  & -7.25 & 24.58246473 & 3.371456537 & 0.0971286497 \tabularnewline
\hline
1 & 1 & -1 & 2.646942620  & -2 & 1.778392210& 0.3275112306$i$  & -0.5-2.065665623$i$\tabularnewline
  & 2 &    & 5.560995211  & -4 & 7.212994786& 1.439683506 & 0.3470628365 \tabularnewline
  & 3 &    & 8.470048808  & -6 & 16.27381376& 2.578075207 & 0.1846098170 \tabularnewline
  & 4 &    & 11.37466883  & -8 & 28.93537848& 3.693450548 & 0.1261531208 \tabularnewline
  & 5 &    & 14.28395410  & -10& 45.26207589& 4.794063129 &0.09615880544 \tabularnewline
\hline
\end{tabular}
\label{T3}
\end{table}

\begin{table}
\caption{The numerical coefficients for the model WFs of the form (\ref{50}), (\ref{51}) and (\ref{57}). Subscript $x$ (of $c_x$) equals to $"+,-"$ and $"0"$ for the representations (\ref{50}), (\ref{51}) and (\ref{57}), respectively.}
\begin{tabular}{|c|c|c|c|c|c|}
\hline
$\lambda$ & $Z$ & $B$ & $\sigma$ & $\kappa$ & $c_{x}$\tabularnewline
\hline
\hline
 0.5 & 1 & 1.519829520 & 0.04865213037 & 9.628915266 & -0.0556618156~~~~  \tabularnewline
-0.5 &   & 1.832098555 & 0.3609211660  & 3.014612921 & -7.507735665~~~~~ \tabularnewline
\hline
   0 & 1 & 1.080682189 & 0.6095047994  & 1.006449560 & -0.05887527389 ~~\tabularnewline
     & 2 & 3.738927217 & 1.0500994024  & 1.678200898 & -0.007532122480~ \tabularnewline
     & 3 & 6.263006695 & 1.5471496894  & 1.806191595 & -0.002802627245~ \tabularnewline
     & 4 & 8.772175623 & 2.0457702001  & 1.861037143 & -0.001402126493~ \tabularnewline
     & 5 & 11.29036042 & 2.5583633585  & 1.892134385 & -0.0008211469023 \tabularnewline
\hline
\end{tabular}
\label{T4}
\end{table}

\end{document}